# Group velocity modulation and wavelength tuning of phonon-polaritons by engineering dielectric-metallic substrates

Flávio H. Feres[1,2*], Rafael A. Mayer[1,3], Ingrid D. Barcelos[1], Raul de O. Freitas[1] and Francisco C. B. Maia[1,*]

*1 – Brazilian Synchrotron Light Laboratory (LNLS), Brazilian Center for Research in Energy and Materials (CNPEM), Zip Code 13083-970, Campinas, Sao Paulo, Brazil.*

*2 - Physics Department, Institute of Geosciences and Exact Sciences, São Paulo State University – UNESP, Rio Claro 13506-900, Brazil*

*3 – Physics Department, Gleb Wataghin Physics Institute, University of Campinas (Unicamp), 13083-859 Campinas, Sao Paulo, Brazil*

\* Corresponding authors: francisco.maia@lnls.br, flavio.feres@lnls.br

### Abstract

In analogy to the observed for single plasmon-polaritons, we show that subdiffractional hyperbolic phonon-polariton ($HP^2$) modes confined in hexagonal boron nitride (hBN) nanocrystals feature wave-particle duality. First, we use Synchrotron Infrared Nanospectroscopy to demonstrate modulation of the $HP^2$ frequency–momentum dispersion relation and group velocity by varying the thickness of the $SiO_2$ layer in the heterostructure $hBN/SiO_2/Au$. These modulations are, then, exploited for the hBN crystal lying on a $SiO_2$ wedge, with a gradient of thickness of such a dielectric medium, built into the Au substrate. Simulations show that a phonon-polariton pulse accelerates as the thickness of the wedge increases. This is explained by a parameter-free semi-classical approach considering the pulse as free quantum particle. Within this picture, an estimated average acceleration value of ~ $1.45 \times 10^{18} \, m.s^{-2}$ is determined using experimental inputs. This estimation is in good agreement with the value of $2.0 \times 10^{18} \, m.s^{-2}$ obtained from the theory directly.

Polaritons [1–3] are *quasi*-particles formed when photons couple to a matter resonance. Therefore, they can intrinsically present the wave-particle duality as reported from single plasmon-polariton – coupling of photons and free electron oscillations - featuring associated quantum and wave phenomena in metallic nano-wires [4,5] and in elaborated plasmonic platforms [6]. Those findings have opened new possibilities for quantum control of light confined at the nanometer scale in the field of quantum plasmonics [7]. Here, we report the wave-particle duality from phonon-polariton waves propagating in van der Waals crystals of hexagonal boron nitride (hBN) crystals lying onto dielectric ($SiO_2$) – metal (Au) metasurface. The wave character manifests itself from the natural interference among polariton waves, whilst the particle aspect emerges from the semi-classical acceleration of the polariton pulse.

The optical near-field properties of hBN is defined in terms of its phononic resonant dielectric tensor $\overleftrightarrow{\varepsilon} = (\varepsilon_{xx}, \varepsilon_{yy}, \varepsilon_{zz})$. This crystal is in-plane isotropric ($\varepsilon_{xx} = \varepsilon_{yy}$) and out-of-plane anisotropic ($\varepsilon_{xx} = \varepsilon_{yy} \neq \varepsilon_{zz}$) featuring type I and type II hyperbolic frequency ($\omega$) – momentum ($q$) dispersion relations. The type I $\omega - q$ occurs at 760 – 820 cm$^{-1}$, wherein $\varepsilon_{zz}$ is resonant with out-of-plane phonons whist $\varepsilon_{xx}$ is a real positive constant, thereby, satisfying the type I hyperbolic condition, $Re(\varepsilon_{zz}) < 0$ and $Re(\varepsilon_{xx}) > 0$. In contrast, the type II $\omega - q$ appears at 1365-1610 cm$^{-1}$, since $\varepsilon_{xx}$ becomes resonant with in-plane phonons, with $\varepsilon_{zz}$ being a positive real constant, leading to type II hyperbolic condition, $Re(\varepsilon_{zz}) > 0$ and $Re(\varepsilon_{xx}) < 0$. In the corresponding infrared windows, therefore, high momenta type I and type II hyperbolic phonon-polaritons (HP$^2$) exist in hBN. The nanooptics of such polaritons has been recently characterized, for a cornucopia of hBN-based systems [8], using high-momenta optical techniques like photo-induced force microscopy (PiFM) [9] [10] and scattering-scanning near-field optical microscopy (s-SNOM) [8] and synchrotron infrared nanospectroscopy (SINS) [11].

Here, we use SINS (Figure 1a), a s-SNOM analogue nanoscopy tool, to image type II HP$^2$ waves in hBN crystals lying on $SiO_2$ thin films, with different thicknesses *d*'s, deposited on Au substrate (Figure 1b and 2). In accordance with theoretical predictions (Figure 1b), we show modulation of the $\omega$-$q$ (Figure 2a-i) of the type II HP$^2$ waves for *d* varying from 0 (direct contact with the metal [8,9]) to values ≥ 200 nm. Accordingly, it is also seen the dependence of the associated group velocity, $v_g$ ($v_g = \frac{d\omega}{dq}$), on *d*. Such relation between $v_g$ and *d* is subsequently exploited by numerical simulations of the dynamics of a polariton pulse in a 80 nm-thick hBN lying on a $SiO_2$ wedge sculpted in an Au substrate (Figure 3). Using the finite-difference time-domain (FDTD) method, we observe a non-linear temporal displacement of the polariton pulse on the wedge in comparison with a linear one on flat Au. The discerning behaviors are explained by treating the polariton as a free quantum particle of effective mass *m\**. Within this approach, we derive a parameter-free semi-classical equation of motion, which matches the numerical simulations (Figure 5) and predicts an average acceleration value of $2.0 \times 10^{18} \ m.s^{-2}$ for the pulse on the wedge. In agreement with those predictions, we find a semi-empirical acceleration value of $1.45 \times 10^{18} \ m.s^{-2}$ by using experimental values of $q$ as inputs for an approximated expression from the referred theory.

In the following, we first present the predicted tunability of $\omega - q$ (Figure 1b) with *d* and the corresponding experimental confirmation from spectral linescans measurements (Figure 2). The consequent dependence of $v_g$ (Figure 3) on *d* is shown and, then, we discuss the acceleration of the polariton pulse on the wedge (Figure 4 and 5).

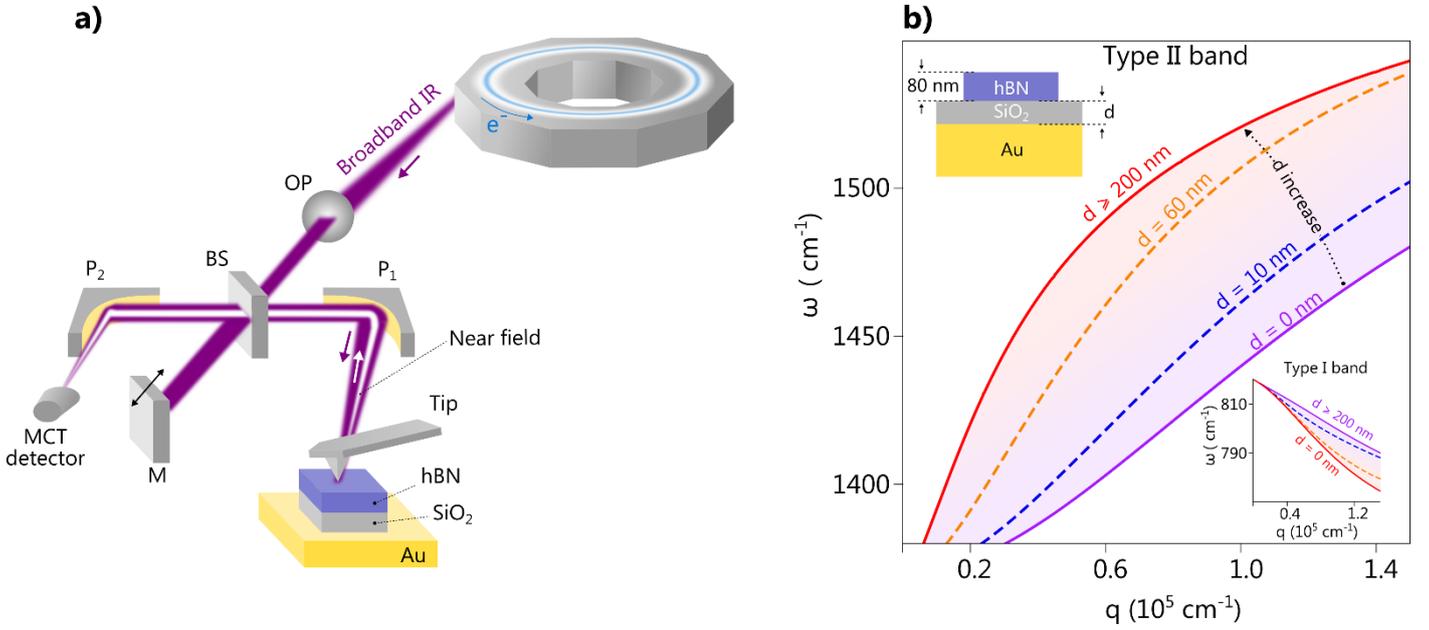

Figure 1. a) Scheme of the experimental setup of SINS: OP – primary optics, BS – beam splitter, M – movable mirror, $P_{1,2}$ – parabolic mirrors. b)Theoretical predictions of the $\omega - q$'s, in the type II band, for the hBN/SiO$_2$/Au metamaterial (upperleft illustration) with $d$ = 0, 10, 60 and > 200 nm. A 80 nm-thick hBN is considered in those calculations. The corresponding analysis, in the type I band, is shown in the downright inset.

Figure 1b) shows the predicted type II $\omega - q$'s for a 80 nm thick hBN on SiO$_2$/Au with different $d$'s. Those calculations are given from the poles of the global reflection coefficient ($r$) of the air/hBN/SiO$_2$/Au stratified medium [12] (see supplemental materials bellow). It is important to comment that exact [9,13] and/or approximated [12] analytical formulae of $\omega - q$ are derivable only for simple cases as hBN/substrate, but not for systems with multiple interfaces like the one above. In such cases, the imaginary part of $r$ has been used as a general method to evidence the $\omega - q$ regions wherein the different orders of polariton modes can exist. Instead of regions, we present numerical curves of $\omega - q$, corresponding to more exact values of the poles of $r$, showing that type II $\omega - q$ can be continuously set for $d$ varying from 0 to 200 nm. Note that, for a given $\omega$, $q$ can be tuned from the highest value at $d$ = 0, where the crystal is in direct contact with the metal, the lowest one at $d$ = 200 nm. From $d$ = 200 nm, $\omega - q$ is unaffected by increasing $d$. In comparison, we note that type I $\omega - q$ presents opposite behavior (Figure 2b inset), $q$ increases as function of $d$ in shorter range yet. The predictions on the type II $\omega - q$ are verified from the analyses of the SINS spectral linescans performed across the edges of ~ 80 nm-thick hBNs on SiO$_2$/Au metasurfaces with $d$ = 0, 10, 60 and 300 nm (Figure 2a-d). In those measurements, polariton modes manifest themselves by contrasting spatio-spectral branches denoted by the arrows in Figure 2a. One can see that, as $d$ increases from 0 to 300 nm, i.e. Figure 2a to Figure 2d, the spatial separation between the maxima enlarges. As shown in the following, this means that $q$ decreases with $d$.

To determine the experimental type II $\omega - q$ for each $d$, the Hertzian dipole antenna (HDA) modelling [14] is used to fit the polariton waves obtained from the SINS spectral linescans (Figure 2a-d) for fixed $\omega$'s along the band. As each linescan is performed across sharp edge of a flat hBN, the tip and the crystal edge are the main polariton launchers treated as dipole source within the HDA modelling. The suitability of the HDA to our case is observed by the agreement of the fit curves with the polariton profiles for $\omega$ = 1445 cm$^{-1}$ for different values of $d$ (Figure 2e-h). Extending this analysis in the entire band, we determine the experimental type II $\omega - q$ for each $d$ matching the corresponding predicted curves in Figure 2i. From the curves in Figure 2i, we find the theoretical (supplemental materials bellow) and experimentally determined $v_g$ as a function of $d$ for a polariton pulse centered at $\omega = 1445$ cm$^{-1}$ (Figure 2h). This figure also exhibits the tunability of $q$ at $\omega = 1445$ cm$^{-1}$ from $1.0 \times 10^5$ cm$^{-1}$ (polariton wavelength $\lambda = 2\pi/q = 0.63$ μm) to $0.34 \times 10^5$ cm$^{-1}$ ($\lambda = 1.8$ μm) for $d$ going from 0 to 300 nm. It has been shown that $q$ can be tuned according to the hBN thickness (*10*) and that the momentum of hybrid hyperbolic plasmon phonon-polaritons in a graphene-hBN device is controllable by gate voltage [12]. We demonstrate, however, $q$ tuning for crystals with nearly the same thicknesses just by altering $d$. An analogous metasurface engineering was employed to tune momenta and phase velocity of graphene plasmon-polaritons [18]. Therefore, the metamaterial assembly [polaritonic media]/[dielectric layer]/[metallic surface] can be a universal scheme allowing tuning the polariton nanophotonic properties with possibility of active control using piezo-electric or phase changeable materials as the intermediate dielectric layer.

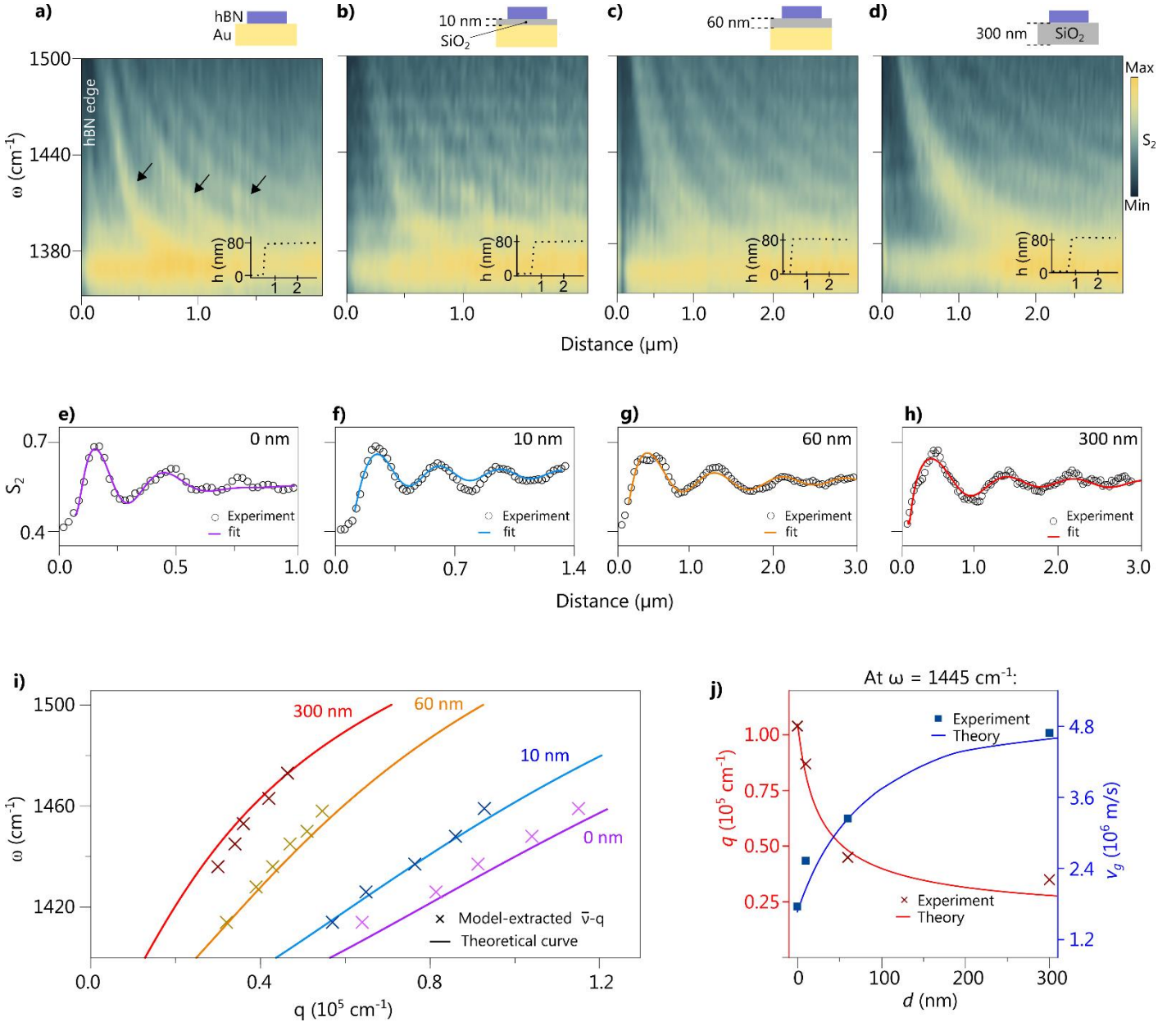

Figure 2. **a-d**. SINS spectral linescans across the edge of hBN crystals of thicknesses around 80 nm lying onto SiO$_2$/Au metasurfaces with $d$ = 0, 10, 60 and 300 nm, respectively. **e-h**. Experimental profiles (open circles) extracted from (a-d) $at\ \omega = 1445\ cm^{-1}$ and the corresponding HDA fits (solid curves). **i**. Fit-determined (symbols) and theoretical $\omega - q$ for each $d$ in (a-d). **j**. Tunability of $q_p$ and $v_g$ as a function of $d$ for $\omega = 1445\ cm^{-1}$. The solid curves are theoretical predictions and symbols are values determined from the HDA fits to experimental polariton profiles.

To exploit this dependence of $v_g$ on d, we examine the propagation of a polariton pulse in a 80 nm-thick hBN onto a wedge of SiO$_2$ built inside the Au substrate (Figure 3a-d) in comparison with a flat Au substrate (Figure 3e-h). The simulations consider a dipole excitation source (D), centered at $\omega_c = 1440$ cm$^{-1}$, with a bandwidth $\Delta\omega = 37\ cm^{-1}$ and located 170 nm above the edge of the crystal. The color scale of Figure 3a corresponds to the magnitude of the electric field z-component wave profile (white line) at $\omega_c$. One can see that such profile continuously increases with $d$ in the whole extension of the wedge. In contrast, the wave profile on the flat Au substrate remains constant for the same propagated distance (Figure 3e). Comparing temporal evolution for three timeframes, we observe that pulse travels faster on the wedge (Figure 3b-d) than on the flat Au substrate (Figure 3f-h). Such simulation results are better visualized from plotting the positions ($x$) of the pulses on the wedge (o) and on the flat Au (×) for a set of progressive timeframes ($t$) in Figure 4a. On the flat Au, $x \times t$ is described by the linear equation of motion $x^{Au}(t) = v_g^{Au} t$ where the constant group velocity $v_g^{Au} = 1.85 \times 10^6\ m/s$ has been derived from the classical electromagnetic theory in Figure 2j. The behavior on wedge, however, is explained by a semi-classical theory. The polariton pulse is regarded a quantum particle with kinetic energy $T$ (Equation 1) and with $m^*$ is given by the second derivative of the resonant polariton energy, $E_{HP^2} = \hbar\omega$, with respect to $q$. Note that on the wedge this expression of $E_{HP^2}$ is a function of $q$ that changes with $x$ even for a fixed $\omega$, as shown in Figure 3a. Equaling the Equation 1 to the classical kinetic energy $T = \frac{m^*}{2}\left(\frac{dx^w}{dt}\right)^2$, we obtain the theoretical equation of motion $x^w(t)$ on the wedge from solving the Equation 3. Such a solution is accomplished since $v_g^w(x)$ and $q_p(x)$ are given by $v_g^w(d)$ and $q_p(d)$ in Figure 2j with the variable substitution $d = x.\tan(\beta)$ where $\beta$ is opening angle of the wedge (Figure 3a). As shown in Figure 4a, the resultant $x^w(t)$ reasonably matches the simulations for the wedge. The straightforwardly calculated group velocity $v_g^w(t) = \frac{dx^w}{dt}$ (Figure 4b) presents a sharp increase from 0 to 2 ps, when $d \sim 60$ nm. Then, it asymptotically tends to the value $\sim 4.0 \times 10^6\ m/s$, when $d = 150$ nm, approaching the value of $4.2 \times 10^6\ m/s$ predicted from Figure 2i, which agrees with the experiment.

In addition, we derive an approximate expression for the average acceleration $\bar{a}_e$ (Equation 4) by assuming that the work realized over the pulse is entirely converted into the variation of its kinetic energy (supplemental materials bellow). In this equation, $\overline{m^*} \sim -2.43 \times 10^{-4}\ m_0$ is the effective mass averaged over the considered interval of $q$ ($m_0$ is the electron mass in vacuum, see supplemental materials below). Negative values of $m^*$ are expected for polaritons [3] when there is a negative curvature of $\omega - q$ as seen for the type II band near low values of $q$. With $q_i = 1.04 \times 10^5$ cm$^{-1}$ at $d = 0$ and $q_f = 0.45 \times 10^5$ cm$^{-1}$ at $d = 60$ nm, it is found $\bar{a}_e \sim 1.45 \times 10^{18}\ m/s^2$ that is in accord with the time-average value of $\bar{a} = 2.\times 10^{18}\ m/s^2$ obtained from the $a(t) = \frac{d^2 x^w}{dt^2}$ curve in Figure 4b.

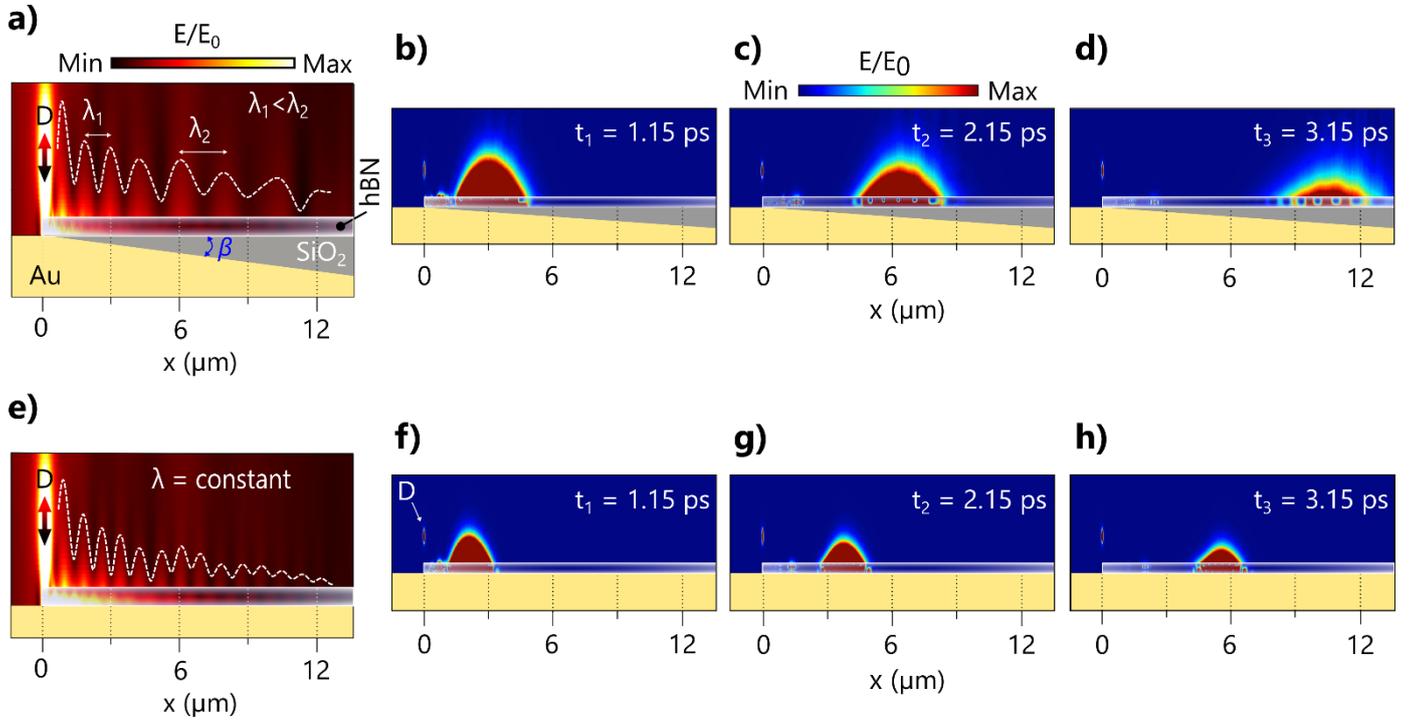

Figure 3. **a)** Simulated profile obtained from Fourier Transform of a type II HP² pulse travelling in a 80 nm-thick hBN on a SiO₂ wedge constructed into Au substrate. $\lambda_{1,2}$ - the wavelength of pulse at distinct positions ($x$) on the wedge. $\beta = 0.76°$ is the wedge opening angle. (**b-d**) values of $x$ of the pulse, on the wedge, at the timeframes $t_{1,2,3}$. **e)** Simulated profile, in analogy with (**a**), of a type II HP² pulse propagating on flat Au. (**f-h**) values of $x$ of the pulse, on the flat Au, at the timeframes $t_{1,2,3}$. In both case, the simulations consider a dipole (D), centered at 1440 cm⁻¹, as the excitation source and a 80 nm thick hBN.

$$T_{HP^2} = \frac{\hbar^2 q^2(\omega_0, x)}{2m^*} \quad (1)$$

$$(m^*)^{-1} = \frac{1}{\hbar^2} \frac{\partial^2 E_{HP^2}}{\partial q_p^2} \quad (2)$$

$$t = \int_0^{x^w(t)} \frac{1}{\left[\frac{\partial v_g^w(\omega_0, x)}{\partial q_p}\right] q_p(\omega_0, x)} dx \quad (3)$$

$$\bar{a}_e = \frac{1}{2}\left(\frac{\hbar}{\overline{m^*}}\right)^2 \frac{(q_f^2 - q_i^2)}{d/\tan(\beta)} \quad (4)$$

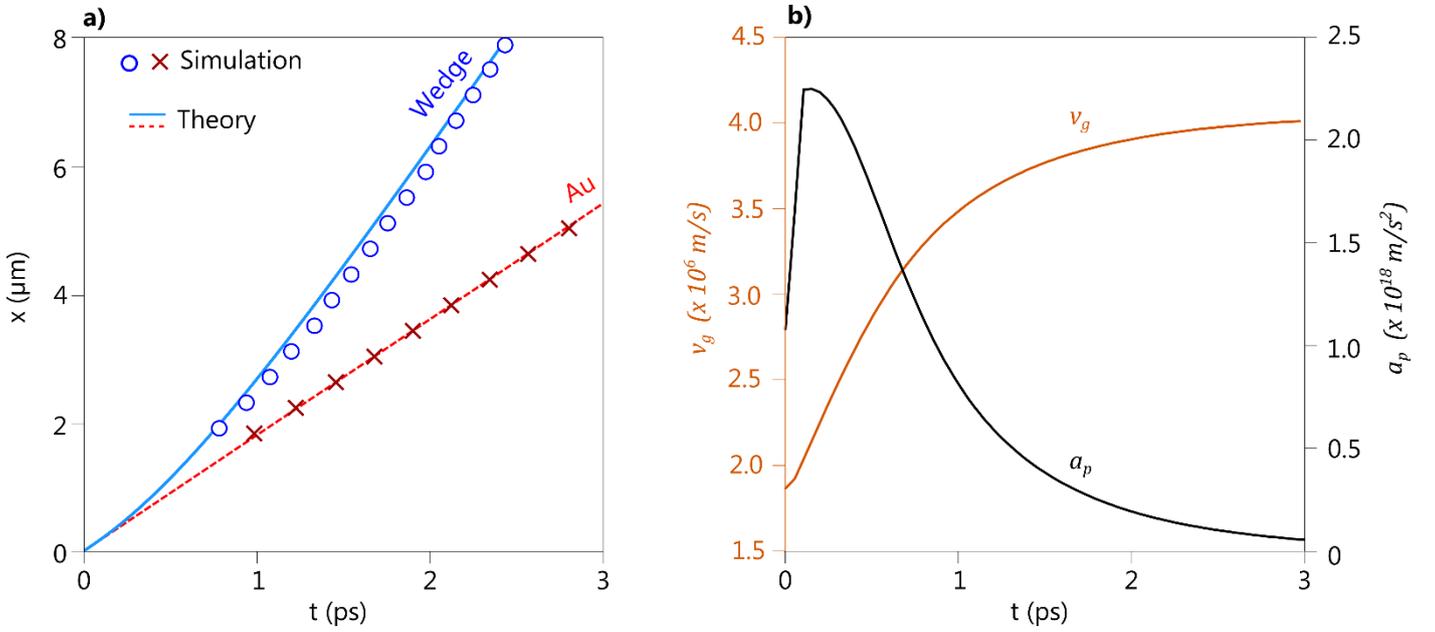

Figure 4 **a)** The red and blue lines are theoretical prediction for motion equation, x ( t ), for HP² pulse in hBN with 80 nm thickness lying on Au and SiO₂ wedge substrates, respectively , whereas the red and blue symbols denote the polariton pulse displacement obtained by the time-resolved simulation. (**b**) Theoretical predictions for group velocity (orange curve) and pulse acceleration (black curve) as function of time, $v_g(t)$ and $a_p(t)$, obtained by the direct first- and second derivative of x ( t ), respectively.

We note that acceleration of exciton-polaritons [19], due to a gradient of thickness in a GaAs microcavity, has been demonstrated. Also, the decrease of $q$ of type I and type II HP² waves with the increase of the hBN thickness was also reported [12,15]. Nevertheless, our findings of the semi-classical acceleration of the subdiffractional type II HP² pulse happens, without any modification of the hBN crystal, but because of use of a dielectric wedge built into a metallic substrate. In the direction that $d$ increases in the wedge, $m^*$ decreases faster than $q$ leading to a liquid augment of $T$ and, accordingly, of $v_g^w$. Importantly, the opposite behavior is expected for type I HP²s. Those results are confirmed from the convergence among experimental, theoretical and simulated analyses. Since there the semi-classical behavior is not a particularity of HP² waves in hBN, it possesses a general aspect that can be exploited in plasmon- and exciton-polaritons in the diverse class of vdW crystal. Hence our findings can be insightful to further investigations using more elaborated experimental apparatus, like pump-probe [20] and cryogenic [21] s-SNOM, which can uncover novel quantum optical phenomena of subdiffractional polaritons.

**Acknowledgements.** We thank the Brazilian Synchrotron Light Laboratory (LNLS) for the beamtime for experiments realizations. Yves Petroff is thanked for the thoroughly revising the work. NeaSpec GmbH is thanked for the technical assistance. Angelo L. Gobbi and Maria H. de Oliveira Piazetta, from the Brazilian Nanotechnology National Laboratory, are thanked for helping the sample construction. F. C. B. M. and F. H. F acknowledge the FAPESP project 2018/05425-3. R. O. F. and R. A. M. acknowledge the FAPESP project 2019/08818-9. R. O. F. acknowledges CNPq for research grant 311564/2018-6.

**Samples construction.** The SiO₂/Au metasurfaces were constructed via deposition, by sputtering, of SiO₂ layers, with thicknesses $d$ = 10 and 60 nm, onto 100 nm thick Au films sputtered on a 10 × 10 × 1 mm³ Si wafer. In sequence, the hBN crystals were mechanically exfoliated (Scotch tape technique) atop each SiO$_2$/Au metasurface. For the hBN/Au ($d = 0$) sample, the crystals were exfoliated directly on the metallic surface of the mentioned Au film. For the sample with $d = 300$ nm, the exfoliation was done directly on a commercial SiO$_2$/Si wafer, with 300 nm-tick oxide layer. The measured crystals in were carefully chosen to have approximately the same thicknesses ~ 80 nm.

**Simulations.** The simulation time was set to 8 ps and the pulse length 0.4 ps. To resolve the SiO$_2$ wedge inclination of $\beta = 0.76°$, the mesh resolution was set to 8 nm horizontally and 2 nm vertically. Optical properties of Au and SiO$_2$ were obtained from Palik [22]. Moreover, Drude-Lorentz model was used to obtain the dielectric function of hBN.

# Supplemental Materials

## 1 Dispersion Relation

The dispersions curves are calculated from the poles of the effective reflectivity coefficient $r_p$ of the air/hBN/SiO$_2$/Au stratified medium (Equation S1). The poles are visualized from plotting the imaginary part of Equation S1 as a map of the excitation frequency $\omega$ versus the real part of the polariton momentum $q$. This equation considers incident p-polarized light and depends on the reflectivity coefficients of the air/hBN interface, $r_a$ (Equation 2), of the hBN/SiO$_2$/Au layered medium, $r_s$ (Equation 3). $k_{ez} = \sqrt{\varepsilon_{xx} k_0^2 - \frac{\varepsilon_{xx}}{\varepsilon_{zz}} q^2}$ is the in-plane momentum of the light that exists hBN. $k_0$ is the momentum of the incident light. $\varepsilon_{xx}$ and $\varepsilon_{zz}$ are the phononic resonant in- and out-of-plane components of the permittivity tensor of the hBN crystal given by Equation S4, with $\rho = xx$ and $zz$. The thickness of the hBN crystal thickness is $d_{hBN}$. $\omega_{TO\rho}$, $\omega_{LO\rho}$ are the transversal and longitudinal optical phonons frequencies, $\varepsilon_{\infty\rho}$ is the asymptotic dielectric value for high frequencies and $\Gamma_\rho$ is the dielectric loss [12,20,23]. In the range of type I band, for out-of-plane resonances, the permittivity constants assume the values of $\omega_{TO\perp} = 750$ cm$^{-1}$, $\omega_{LO\perp} = 820$ cm$^{-1}$, $\varepsilon_{\infty\perp} = 4.9$ and $\Gamma_\perp = 5$ cm$^{-1}$. For in-plane resonance frequencies, type II band, was used the values of $\omega_{TO\perp} = 1370$ cm$^{-1}$, $\omega_{LO\perp} = 1610$ cm$^{-1}$, $\varepsilon_{\infty\perp} = 4.9$ and $\Gamma_\perp = 5$ cm$^{-1}$.

$$r_p = \frac{r_a - r_s e^{i2k_{ez}d_{hBN}}}{1 + r_a r_s e^{i2k_{ez}d_{hBN}}} \qquad \text{Equation S1}$$

$$r_a = \frac{\epsilon_\perp k_{air} - \epsilon_{air} k_{ez}}{\epsilon_\perp k_{air} + \epsilon_{air} k_{ez}} \qquad \text{Equation S2}$$

$$r_s = \frac{r_a' - r_s' e^{i2k_{SiO_2}d}}{1 + r_a' r_s' e^{i2k_{SiO_2}d}} \qquad \text{Equation S3}$$

$$\varepsilon_\rho = \varepsilon_{\infty\rho}\left(1 + \frac{(\omega_{LO}^2)_\rho - (\omega_{TO}^2)_\rho}{(\omega_{TO}^2)_\rho - \omega^2 - i\omega\Gamma_\rho}\right) \qquad \text{Equation S4}$$

Equation 3 is expressed by $r_a'$ (Equation S5) and $r_s'$ (Equation S6) that are the reflection coefficients of the hBN/SiO$_2$ and SiO$_2$/Au interfaces, respectively. Equations S2 – S5 take into account the permittivities of the air $\varepsilon_{air} = 1$, of the SiO$_2$ layer $\varepsilon_{SiO_2}$ (Equation S6) and of the Au substrate $\varepsilon_{Au}$ (Equation S7). They also regard the z-axis momentum $k_i = \sqrt{\varepsilon_i k_0^2 - q_p^2}$, for each media, where $i$ = air, Au and SiO$_2$. The thickness of the SiO$_2$ layer is $d$, as defined in the main manuscript.

$$r_a' = \frac{\varepsilon_{SiO_2} k_{ez} - \varepsilon_\perp k_{SiO_2}}{\varepsilon_{SiO_2} k_{ez} + \varepsilon_\perp k_{SiO_2}} \quad \text{Equation S4}$$

$$r_s' = \frac{\varepsilon_{Au} k_{SiO_2} - \varepsilon_{SiO_2} k_{Au}}{\varepsilon_{Au} k_{SiO_2} + \varepsilon_{SiO_2} k_{Au}} \quad \text{Equation S5}$$

$$\varepsilon_{SiO_2} = \varepsilon_\infty \left(1 + \sum_j \frac{(\omega_{LO}^2)_j - (\omega_{TO}^2)_j}{(\omega_{TO}^2)_j - \omega^2 - i\omega \Gamma_j}\right) \quad \text{Equation S6}$$

$$\varepsilon_{Au} = 1 - \frac{\omega_p^2}{\omega^2 + i\omega \Gamma_{Au}} \quad \text{Equation S7}$$

Equation S7, for $\varepsilon_{Au}$ consists of the Drude Model, where $\omega_p$ and $\Gamma_{Au}$ are the plasmonic frequency and plasmonic damping, respectively [24]. Moreover, $SiO_2$ permittivity, is described by multiple Lorentz-Drude due to the optical surface phonon polariton resonance centered near ω = 1128 cm$^{-1}$ and the amorphous morphology of sputtering $SiO_2$, here $\omega_{TO}$, $\omega_{LO}$ and $\Gamma_{SiO_2}$ are the transverse optical phonon frequency, longitudinal optical phonon frequency and the crystal dielectric loss, respectively [25], [26].

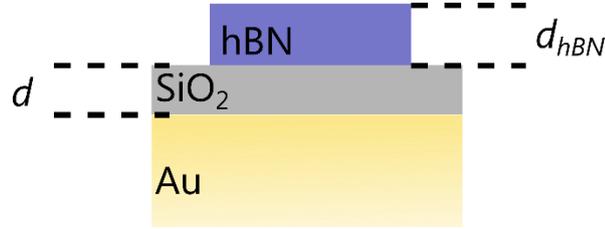

Figure S1 - Schematic representation of hBN with thickness *d*<sub>hBN</sub> laying on metasurface composed by a SiO2 amorphous film with *d* thickness and Gold layer

## 2 Momentum and Group Velocity as Function of SiO₂ thickness for Type II Band

Momentum of the HP² extracted by the fittings, at 1436 and 1460 cm$^{-1}$, has good agreement with the theoretical predictions as shown in Figure S2. The group velocity is given by $v_g = \frac{\partial \omega}{\partial q}$ and is shown in Figure S3.

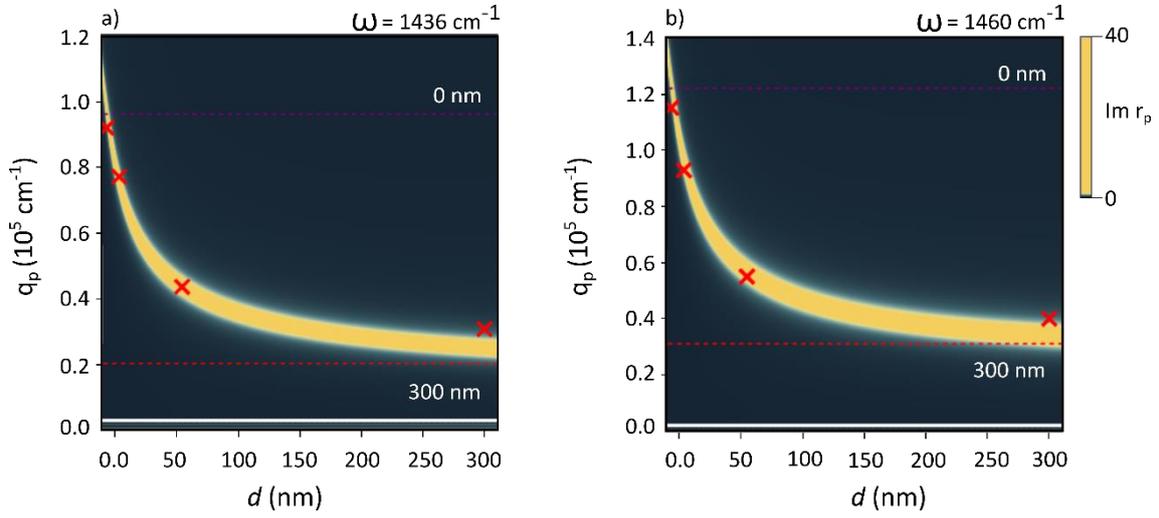

Figure S2 - Theoretical momentum as function of SiO2 thickness, obtained by the false color plot of *Im* rp (Equation S1) and experimental momentum extracted by fittings (red symbols) for **a)** $\omega = 1436$ cm$^{-1}$ and **b)** $\omega = 1460$ cm$^{-1}$

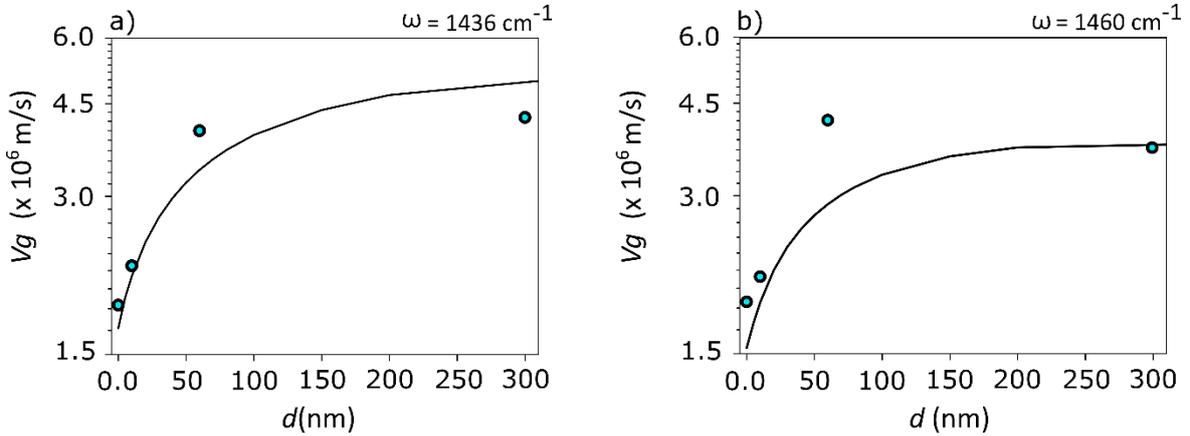

Figure S3- Theoretical (black solid lines) and experimental (green symbols) group velocities as function of SiO$_2$ thickness for a) for **a)** $\omega = 1436$ cm$^{-1}$ and **b)** $\omega = 1460$ cm$^{-1}$

## 3 Type I band Tuning

In the main text we concern to describe and show the HP² subwavelength tuning for type II hyperbolic band of hBN 80 nm crystal. Figure S4 shows the type I linescans results. Here we cannot see the HP² standing waves profiles clearly due the spectral resolution used to define primarily type II band. However, we can see a tiny blueshift of the first maximum (inverse presented in type II), as expected theoretically, as show in inset Figure 1a in the main text and with more details in Figure S5. The type I tuning occurs more abruptly than type II, where in almost all frequencies, when $d \geq 75$ nm, the momentum assumes constant values.

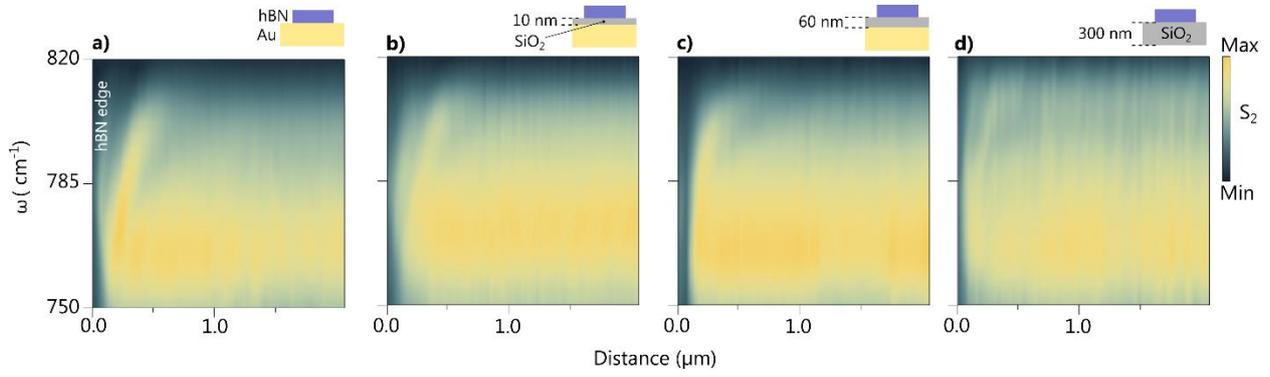

Figure S4-. SINS spectral linescans for type I band (750 – 820 cm$^{-1}$) across the edge of hBN crystals of thicknesses around 80 nm lying onto SiO$_2$/Au metasurfaces with d = **a)** 0, **b)** 10, **c)** 60 and **d)** 300 nm, respectively.

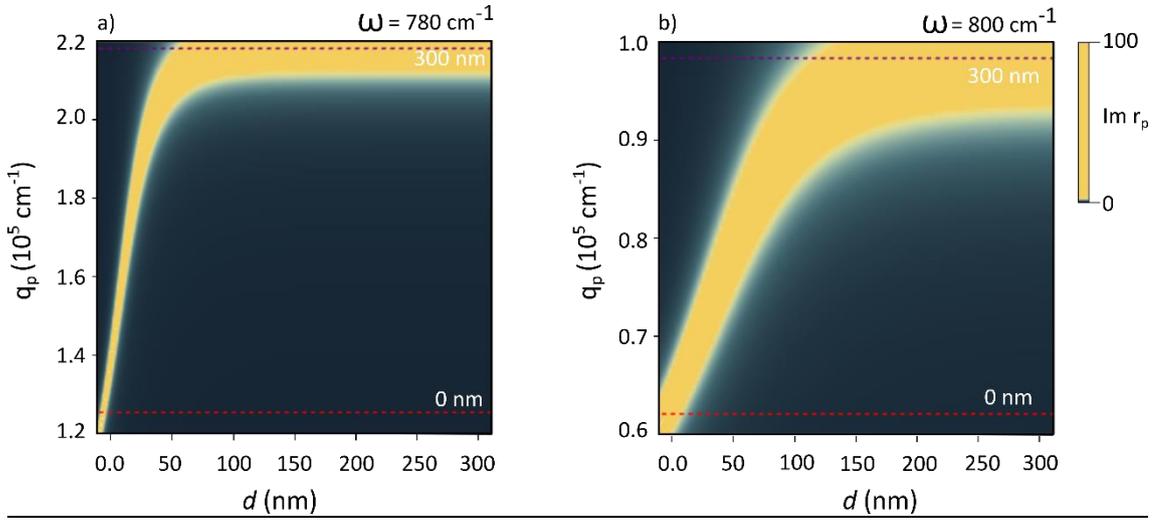

Figure S5-Theoretical momentum as function of SiO$_2$ thickness, obtained by the false color plot of *Im* rp (Equation S1) for **a)** $\omega$ = 780 cm-1 and **b)** $\omega$ = 900 cm$^{-1}$

## 4 Negative effective mass

The effective mass of the polariton is determined as $(m^*)^{-1} = \frac{1}{\hbar^2} \frac{\partial^2 E_{HP^2}}{\partial q_p^2}\Big|_{q=0}$ and can assume positive or negative values [3,27]. For type II hyperbolic polaritons at the resonance for $q = 0$, we find $m_p \sim -2.43 \times 10^{-4} \, m_0$ where $m_0 = 9.1 \times 10^{-31} \, kg$ is the vacuum electron inertial mass. Figure S6 a) shows the dispersion relation for type II band of *hBN* (80 nm thick) on *Au* and Figure S6 b) shows the effective mass curve as function of in-plane polariton momentum.

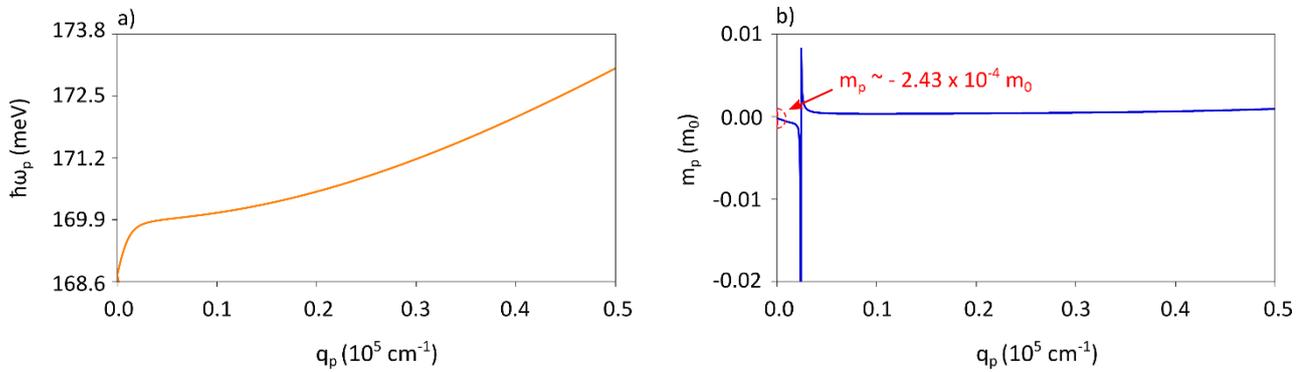

Figure S6-Type II dispersion **a)** and polariton effective mass **b)**

## 5 References of the Supplemental Materials